\documentstyle[twocolumn,aps,epsf]{revtex}

\begin{document}
\draft
\preprint{HEP/123-qed}
\title{Quantum Pattern Recognition}
\author{C. A. Trugenberger}
\address{InfoCodex SA, av. Louis-Casai 18, CH-1209 Geneva, Switzerland\\
Theory Division, CERN, CH-1211 Geneva 23, Switzerland}
\address{e-mail: ca.trugenberger@InfoCodex.com}
\date{\today}
\maketitle
\begin{abstract}
I review and expand the model of {\it quantum associative memory} that I have
recently proposed. In this model binary patterns of $n$ bits
are stored in the quantum superposition of the appropriate subset 
of the computational basis of $n$ qbits. Information can be retrieved by
performing an input-dependent rotation of the memory quantum state within this
subset and measuring the resulting state. 
The amplitudes of this rotated memory state are peaked on those
stored patterns which are closest in Hamming distance to the input, resulting
in a high probability of measuring a memory pattern very similar to it. 
The accuracy of pattern recall can be tuned by adjusting a parameter playing the
role of an effective temperature. This model solves the well-known capacity
shortage problem of classical associative memories, providing 
an {\it exponential improvement} in capacity. The price to pay is the
probabilistic nature of information retrieval, a feature that, however, 
this model shares with our own brain. 
\end{abstract}

\narrowtext
\section{Introduction}
The power of quantum computation \cite{review} is mostly associated with the
speed-up in computing time it can provide with respect to its classical
counterparts, the paramount examples being Shor's factoring 
algorithm \cite{shor} and Grover's search algorithm \cite{grover}.
There is, however, another aspect of quantum computation which represents a big
improvement upon its classical counterpart \cite{myself}. This leads to an
exponential increase in a particular memory capacity rather than speed. In this
paper I will review and expand
the main aspects of this new application of quantum
information theory. Further aspects of it can be found in \cite{csj}. 

In traditional computers the storage
of information requires setting up a lookup table (RAM). The main disadvantage
of this address-oriented memory system lies in its rigidity. Retrieval of
information requires a precise knowledge of the memory address and, therefore,
incomplete or corrupted inputs are not permitted.

This is definitely not how our own brain works. When trying to recognize a
person from a blurred photo it is totally useless to know that it is the 17384th
person you met in your life. Rather, the recognition process is based on our
strong power of association with stored memories that resemble the given
picture. Association is what we use every time we solve a crossword puzzle and
is distinctive of the human brain.

Given the superior power of associative pattern recognition for complex tasks,
the shortcomings of RAM memories were addressed by introducing 
models of associative (or
content-addressable) memories \cite{neuralnetworks}. Here,
recall of information is possible on the basis of partial knowledge of their
content, without knowing the storage location. These are examples of collective
computation on neural networks \cite{neuralnetworks}, the best known example
being the Hopfield model \cite{hopfield} and its generalization to a
bidirectional associative memory \cite{kosko}. 

While these models solve the problem of recalling incomplete or noisy inputs,
they suffer from a severe capacity shortage. Due to the phenomenon of crosstalk,
which is essentially a manifestation of the spin glass transition \cite{parisi}
in the corresponding spin systems, the maximum number of binary patterns that
can be stored in a Hopfield network of $n$ neurons is $p_{max} \simeq 0.14 \ n$
\cite{neuralnetworks} . While various possible improvements can be introduced
\cite{neuralnetworks}, the maximum number of patterns remains linear in the
number of neurons, $p_{max} = O(n)$.

Quantum mechanics offers a way out from the impossibility of reconciling
the association power of content-addressable memories with the requirement of
large storage capacity.
Indeed, quantum mechanical entanglement provides a natural
mechanism for both improving dramatically the storage capacity of associative
memories and retrieving corrupted or incomplete information. 

The basic idea is to store the given $p$ binary patterns of $n$ bits in a
quantum superposition of the corresponding subset of the computational basis of
$n$ qbits. The number of
binary patterns that can be stored in 
such a {\it quantum associative memory} is exponential
in the number $n$ of qbits, $p_{max} = 2^n$, i.e. it is optimal in the sense
that all binary patterns that can be formed with $n$ bits can be stored. 

The basic idea of the information retrieval mechanism is very simple. Given an
input pattern, the memory quantum state is rotated within the subspace defined
by the stored patterns so that the resulting amplitudes are peaked on the stored
patterns which are closest in Hamming distance to the input. A measurement of
the rotated memory quantum state provides the output pattern. 

An efficient way to perform this rotation is to embed the memory quantum state
in a larger Hilbert space by adding $b$ control qbits. The full state is then
rotated in the enlarged Hilbert space. After this rotation one is interested
only in the projection of the rotated state onto a specific subspace of the
enlarged Hilbert space. This projection can be obtained either by repeated
measurement or by rotating the state (approximately) to the desired subspace
using the amplitude amplification technique \cite{amam}. Either way one has to
repeat a certain algorithm a number of times and measure the control register to
check if the desired projection has been obtained. 
The information retrieval mechanism is thus probabilistic, with postselection 
of the measurement result. 
This means that one has to repeat an algorithm until 
a threshold $T$ is reached or the
measurement of a control register yields a given result. 
In the former case the input is not recognized. In the latter case, instead, 
the output is determined itself by a probability distribution 
on the memory which is peaked around the stored 
patterns closest in Hamming distance to the input.

The accuracy of this information retrieval mechanism depends on the
distribution of the stored patterns. Recognition efficiency is best when the
number of stored patterns is very large while identification efficiency
is best for isolated patterns which are very different from all other ones,
both very intuitive features. Both efficiencies can be tuned to prescribed
accuracy levels. The recognition efficiency can be varied by changing the
threshold $T$: the higher $T$, the larger the number of qbits that can be
corrupted without affecting recognition. The identification efficiency, instead,
can be tuned by varying the number $b$ of control qbits in the memory.
As we shall see, $b=1/t$ plays the role of an inverse effective temperature $t$.
The lower $t$, the more concentrated is the corresponding effective Boltzmann
distribution on the states closest (in Hamming distance) to the input and the
better becomes the identification.

By averaging over the distribution of stored patterns one can eliminate the
dependence on the stored pattern distribution and derive the
effective statistical mechanics of quantum associative memories by introducing
the usual thermodynamic potentials. 
In particular, the free energy $F(t)$ describes the average
behaviour of the recall mechanism at temperature $t$ and provides
concrete criteria to tune the accuracy of the associative memory. By
increasing $b$ (lowering $t$), the associative memory undergoes a phase
transition from a disordered phase with no correlation between input and output
to an ordered phase with minimal Hamming distance bewteen the input and the
output. This extends to quantum information theory the relation with Ising spin
systems known in error-correcting codes \cite{errcorr} and in public key
cryptography \cite{crypto}.

\section{Storing Information}
Let me start by describing the elementary quantum gates \cite{review}
that I will use in the rest of the paper. First of all there are the
single-qbit gates represented by the Pauli matrices $\sigma_i$, $i=1\dots 3$.
The first Pauli matrix $\sigma_1$, in particular, implements the NOT gate.
Another single-qbit gate is the Hadamard gate H,
with the matrix representation
\begin{equation}
H = {1\over \sqrt{2}} \ \left( \matrix{1 & 1\cr
1 & -1\cr} \right) \ .
\label{adda}
\end{equation}
Then, I will use extensively the two-qbit XOR (exclusive OR) gate, which
performs a NOT on the second qbit if and only if the first one is in state
$|1\rangle$. In matrix notation this gate is represented as
${\rm XOR} = {\rm diag} \left( 1, \sigma_1 \right)$, where $1$ denotes a
two-dimensional identity matrix and $\sigma _1$ acts on the components
$|01\rangle$ and $|11\rangle$ of the Hilbert space. The 2XOR, or Toffoli gate
is the three qbit generalization of the XOR gate: it performs a
NOT on the third qbit if and only if the first two are both in state
$|1\rangle$. In matrix notation it is given by 
${\rm 2XOR} = {\rm diag} \left( 1, 1, \sigma_1 \right)$. In the storage
algorithm I shall make use also of the nXOR generalization of these
gates, in which there are n control qbits. This gate is also used in
the subroutines implementing the oracles underlying Grover's algorithm
\cite{review} and can be realized using unitary maps affecting only few
qbits at a time \cite{gates}, which makes it feasible. 
All these are standard gates. In addition to them I introduce the two-qbit
controlled gates
\begin{eqnarray}
CS^i &&= |0\rangle \langle 0| \otimes 1 + |1\rangle \langle 1|
\otimes S^i \ ,
\nonumber \\
S^i &&= \left( \matrix{\sqrt{i-1\over i}&1\over \sqrt{i}\cr
-1\over{\sqrt{i}}&\sqrt{i-1\over i}\cr} \right) \ ,
\label{addb}
\end{eqnarray}
for $i=1, \dots, p$. These have the matrix notation $CS^i = {\rm diag}
\left( 1, S^i \right)$. For all these gates I shall indicate by subscripts
the qbits on which they are applied, the control qbits coming always first.  

Given $p$ binary patterns $p^i$ of length $n$, it is not difficult to imagine
how a quantum memory can store them. Indeed, such a memory is naturally 
provided by the following superposition of $n$ entangled qbits:
\begin{equation}
|m\rangle = {1\over \sqrt{p}}\ \sum_{i=1}^p \ |p^i\rangle \ .
\label{a}
\end{equation}
The only real question is how to generate this state unitarily from a 
simple initial state of $n$ qbits. In \cite{myself} I presented an algorithm
which achieves this by loading sequentially the classical patterns into an
auxiliary quantum register from which they are then copied into the actual
memory register. Here I will generalize this algorithm by constructing also the
unitary operator which generates the memory state (\ref{a}) directly from the
state $|0,\dots, 0\rangle$. 

Let me begin by reviewing the sequential algorithm of ref. \cite{myself}. 
I shall use three registers: a first register $p$ of
$n$ qbits in which I will subsequently feed the patterns $p^i$ to be stored, a
utility register $u$ of two qbits prepared in state $|01\rangle$, 
and another register $m$
of $n$ qbits to hold the memory. This latter will be initially prepared in state
$|0_1, \dots, 0_n\rangle$. The full initial quantum state is thus
\begin{equation}
|\psi _0^1\rangle = |p^1_1, \dots p^1_n; 01; 0_1, \dots, 0_n\rangle \ .
\label{c}
\end{equation}
The idea of the storage algorithm is to separate this state 
into two terms, one corresponding to the already stored patterns, and another
ready to process a new pattern. These two parts will be distinguished by the
state of the second utility qbit $u_2$: $|0\rangle$ for the stored patterns 
and $|1\rangle$ for the processing term. 

For each pattern $p^i$ to be stored 
one has to perform the operations described below:
\begin{equation}
|\psi _1^i \rangle = \prod_{j=1}^n 
\ 2XOR_{p_j^i u_2 m_j} \ |\psi _0^i\rangle \ .
\label{addc}
\end{equation}
This simply copies pattern $p^i$ into the memory register of the processing
term, identified by $|u_2\rangle = |1\rangle$.
\begin{eqnarray}
|\psi _2^i \rangle &&= \prod_{j=1}^n 
\ NOT_{m_j} \ XOR_{p_j^i m_j} \ |\psi _1^i\rangle \ ,
\nonumber \\
|\psi _3^i \rangle &&= nXOR_{m_1 \dots m_n u_1} |\psi_2^i \rangle \ .
\label{addd}
\end{eqnarray}
The first of these operations makes all qbits of the memory register
$|1\rangle$'s when the contents of the pattern and memory registers are
identical, which is exactly the case only for the processing term. Together,
these two operations change the first utility qbit $u_1$ of the processing
term to a $|1\rangle$, leaving it unchanged for the stored patterns term.
\begin{equation}
|\psi_4^i\rangle = CS^{p+1-i}_{u_1 u_2} \ |\psi _3^i\rangle \ .
\label{adde}
\end{equation}
This is the central operation of the storing algorithm. It separates out the
new pattern to be stored, already with the correct normalization factor.
\begin{eqnarray}
|\psi _5^i \rangle &&= nXOR_{m_1 \dots m_n u_1} |\psi_4^i \rangle \ ,
\nonumber \\
|\psi _6^i \rangle &&= \prod_{j=n}^1
\ XOR_{p_j^i m_j} \ NOT_{m_j}\  |\psi _5^i\rangle \ .
\label{addf}
\end{eqnarray}
These two operations are the inverse of eqs.(\ref{addd}) and restore the
utility qbit $u_1$ and the memory register $m$ to their original values. 
After these operations on has
\begin{equation}
|\psi _6^i \rangle = 
{1\over \sqrt{p}}\ \sum_{k=1}^i |p^i;00;p^k\rangle + \sqrt{p-i\over p}
|p^i;01;p^i\rangle \ .
\label{d}
\end{equation}
With the last operation,
\begin{equation}
|\psi_7^i\rangle = \prod_{j=n}^1\ 2XOR_{p^i_j u_2 m_j}
\ |\psi_6^i\rangle \ ,
\label{addg}
\end{equation}
one restores the third register $m$ of the processing term, the second term
in eq.(\ref{d}) above, to its initial value $|0_1, \dots, 0_n\rangle$. At this
point one can load a new pattern into register $p$ and go through the 
same routine as just described. At the end of the whole process, 
the $m$-register is exactly in state $|m\rangle$, eq. (\ref{a}).

Any quantum state can be generically obtained by a unitary transformation of the
initial state $|0,\dots ,0\rangle$. This is true also for the memory state
$|m\rangle$. In the following I will explicitly construct the unitary operator
$M$ which implements the transformation $|m\rangle = M \ |0,\dots ,0\rangle$.

To this end I introduce first the single-qbit unitary gates 
\begin{equation}
U^i_j = {\rm cos} \left( {\pi \over 2} p^i_j \right) 1 + i \ {\rm sin} \left(
{\pi \over 2} p^i_j \right) \sigma _2 \ ,
\label{newa}
\end{equation}
where $\sigma_2$ is the second Pauli matrix. These operators are such that their
product over the $n$ qbits generates pattern $p^i$ out of $|0,\dots, 0\rangle$:
\begin{eqnarray}
|p^i\rangle &&= P^i \ |0,\dots ,0\rangle \ ,
\nonumber \\
P^i &&\equiv \prod_{j=1}^n U^i_j \ .
\label{newb}
\end{eqnarray}

I now introduce, in addition to the memory register proper, the same two utility
qbits as before, also initially in the state $|0\rangle$. The idea is, exactly
as in the sequential algorithm, to split the state into two parts, a storage
term with $|u_2\rangle = |0\rangle$ and a processing term with
$|u_2\rangle = |1\rangle$. Therefore I generalize the operators $P^i$ defined
above to
\begin{equation}
CP^i_{u_2} \equiv \prod_{j=1}^n \ CU^i_{u_2 j} \ ,
\label{newc}
\end{equation}
which loads pattern $p^i$ into the memory register only for the processing term.
It is then easy to check that
\begin{eqnarray}
&&|m;00\rangle = M \ |0,\dots, 0;00\rangle \ ,
\nonumber \\
&&M = \prod_{i=1}^p \left[ \left( CP^i_{u_2} \right) ^{-1} NOT_{u_1}
CS^{p+1-i}_{u_1 u_2} XOR_{u_2 u_1} CP^i_{u_2} \right] \times 
\nonumber \\
&&\times \ NOT_{u_2} \ .
\label{newd}
\end{eqnarray}
The storage algorithm is thus efficient in the sense that the number
$p(2n+3) + 1$ of elementary one- and two-qbit gates needed to implement it, is
linear in $n$ for fixed $p$. 
Note that, by construction, there are no restrictions on the loading factor
$p/n$. However, the storage algorithm is efficient in an absolute sense only for
$p$ polynomial in $n$. 

\section{Remembering}
A memory is of real value only if it can be used repeatedly.
This poses a problem since, as we shall see in the next section, an output of
the memory is obtained by measuring the memory register and
the rules of quantum mechanics
imply that, when the memory register is measured, all the
information about the entangled superposition of stored patterns is lost. If one
does not want to forget everything after the first information
retrieval one must therefore find a way to store the information for repeated
use. 

In quantum mechanics there are many choices to do this, since information can be
stored, with various degrees of compression, both in quantum states and in
unitary operators. The most compressed storage would be a quantum state: in this
case up to $2^n$ patterns can be stored using only $n$ (quantum) degrees of
freedom. To this end, however, one would have to  
keep a master copy of the memory and produce copies out of it when needed.
Unfortunately, this is impossible since the linearity of quantum mechanics
forbids exact universal cloning of quantum states \cite{zurek}. 
Universal cloning
\cite{unclo} has two disadvantages: first of all the copies to be used for
information retrieval are imperfect, though optimal \cite{opt}; secondly, the
quality of the master copy decreases with each recall, i.e. the memory is
quickly washed out. 

This leaves state-dependent cloning as the only viable option if one wants to
store at least part of the information in a quantum state.  
State-dependent
cloners are designed to reproduce only a finite number of states and this is
definitely enough for our purposes. The simplest option in this setting 
is to use
a probabilistic cloning machine \cite{cloning}. To this end it is sufficient to
consider any dummy state $|d\rangle$ different from $|m\rangle$ (for more than
two states the condition would be linear independence) and to construct a
probabilistic cloning machine for these two states. This machine reproduces
$|m\rangle$ with probability $p_m$ and $|d\rangle$ with probability $p_d$;
a flag tells us exactly when the desired state $|m\rangle$ has been
obtained. In order to obtain an exact copy of $|m\rangle$ one needs then $1/p_m$
trials on average. The master copy is exactly preserved. 

The cloning efficiencies of the probabilistic cloner of two states are bounded
as follows \cite{cloning}:
\begin{equation}
p_m + p_d \le {2\over 1+ \langle d|m \rangle } \ .
\label{neweqa}
\end{equation}
This bound can be made large by choosing $|d\rangle$ as nearly orthogonal to
$|m\rangle$ as possible. A simple way to achieve this for a large number of
patterns is to encode also the state
\begin{equation}
|d\rangle = {1\over \sqrt{p}} \ \sum_{i=1}^p (-1)^{i+1} |p^i \rangle 
\label{neweqb}
\end{equation}
together with $|m\rangle$ when storing information. This can be done simply by
using alternately the operators $S^i$ and $\left( S^i \right) ^{-1}$ in the
storing algorithm of section 2. For binary patterns which are all different from
one another one has then
\begin{eqnarray}
\langle d|m\rangle &&= 0 \ , \qquad \qquad p \ {\rm even} \ ,\\
\nonumber
\langle d|m\rangle &&= {1\over p} \ , \qquad \qquad p \ {\rm odd} \ ,
\label{neweqc}
\end{eqnarray}
and the bound for the cloning efficiencies is very close to its maximal value 2
in both cases. 

The quantum network for the probabilistic cloner of two 
states has been developed in \cite{clonet}. It can be constructed exclusively
out of the two simple distinguishability tranfer (D) and state separation (S)
gates. Note that these gates embody information about the two states to be
cloned. Part of the memory, therefore, actually resides in the cloning network,
which is unavoidable if one wants to use a quantum memory repeatedly. 
This is then a
mixed solution in which part of the information resides in a quantum state and
another part in a unitary operator implemented as a probabilistic cloning
network.  

At the other end of the spectrum one can store the information about the $p$
patterns entirely in the unitary operator $M$ in eq.(\ref{newd}). Each time one
needs to retrieve information one prepares then an initial quantum state 
$|0,\dots,0\rangle$ and one transforms it into $|m\rangle$ by applying $M$ to
it. In this case one needs $pn$ bits of information to store the $p$ patterns,
exactly as in the classical Hopfield model. Indeed, this way of storing the
patterns is very close in spirit to the Hopfield model since a unitary operator
can always be represented as the exponential of a Hamiltonian operator, which is
the quantum generalization of an energy functional. As I now show, however, in
the quantum case there are no restrictions on the number of patterns that can be
stored and retrieved. 

\section{Retrieving Information}
Assume now one is given a binary input $i$, which might be, e.g. a corrupted
version of one of the patterns stored in the memory.
The retrieval algorithm requires also three registers. The first register
$i$ of n qbits contains the input pattern; 
the second register $m$, also of n qbits,
contains the memory $|m\rangle$; finally there is a control
register $c$ with $b$ qbits all initialized in the state $|0\rangle$.

The full initial quantum state is thus:
\begin{equation}
|\psi_0\rangle = {1\over \sqrt{p}} \sum_{k=1}^p |i; p^k;
0_1,\dots , 0_b\rangle
\label{ab}
\end{equation}
where $|i\rangle = |i_1,\dots ,i_n\rangle$ denotes 
the input qbits, the second register, $m$, contains the memory (\ref{a})
and all $b$ control qbits are in
state $|0\rangle$. Applying the Hadamard gate 
to the first control qbit one obtains
\begin{eqnarray}
|\psi _1\rangle &&= {1\over \sqrt{2p}} \ \sum_{k=1}^p |i; 
p^k; 0_1,\dots ,0_b\rangle
\nonumber \\ 
&&+{1\over \sqrt{2p}} \ \sum_{k=1}^p |i; 
p^k; 1_1,\dots ,0_b\rangle \ .
\label{ad}
\end{eqnarray}
I now apply to this state the following combination of quantum gates:
\begin{equation}
|\psi _2\rangle = \prod_{j=1}^n \ NOT_{m_j} 
\ XOR_{i_j m_j} |\psi _1\rangle \ ,
\label{ae}
\end{equation}
As a result of the above operation the memory register qbits
are in state $|1\rangle$ if $i_j$ and $p^k_j$ are identical
and $|0\rangle$ otherwise:
\begin{eqnarray}
|\psi _2\rangle &&= {1\over \sqrt{2p}} \ \sum_{k=1}^p |i; 
d^k; 0_1,\dots ,0_b\rangle 
\nonumber \\
&&+{1\over \sqrt{2p}} \ \sum_{k=1}^p |i; 
d^k; 1_1,\dots ,0_b\rangle \ ,
\label{af}
\end{eqnarray}
where $d^k_j = 1$ if and only if  $i_j=p^k_j$ and $d^k_j=0$ otherwise.

Consider now the following Hamiltonian:
\begin{eqnarray}
{\cal H} &&= \left( d_H \right)_m \otimes \left( \sigma_3 \right)_{c_1} \ ,
\nonumber \\
\left( d_H \right)_m && = \sum_{j=1}^n 
\left( {\sigma_3 + 1\over 2} \right) _{m_j}\ ,
\label{ag}
\end{eqnarray}
where $\sigma _3$ is the third Pauli matrix.
${\cal H}$ measures the number of 0's in register $m$, with a plus sign if $c_1$
is in state $|0\rangle$ and a minus sign if $c_1$ is in state $|1\rangle$. Given
how I have prepared the state $|\psi _2\rangle$, this is nothing else than the
number of qbits which are different in the input and memory registers $i$ and
$m$. This quantity is called the {\it Hamming distance} and represents the
(squared) Euclidean distance between two binary patterns. 

Every term in the superposition (\ref{af}) is an eigenstate of ${\cal H}$ with a
different eigenvalue. Applying thus the unitary operator ${\rm exp} 
(i \pi {\cal H}/2n)$ to $|\psi _2\rangle$ one obtains
\begin{eqnarray}
|\psi _3\rangle &&= {\rm e}^{i{\pi \over 2n}{\cal H}} \ |\psi_2\rangle \ ,
\label{ah} \\
|\psi_3\rangle &&= {1\over \sqrt{2p}} \sum_{k=1}^p {\rm e}^{i{\pi\over 2n}
d_H\left( i, p^k\right)}
|i; d^k; 0_1,\dots ,0_b\rangle 
\nonumber \\
&&+ {1\over \sqrt{2p}} \sum_{k=1}^p {\rm e}^{-i{\pi\over 2n}
d_H\left( i, p^k\right)}
|i; d^k; 1_1,\dots ,0_b\rangle \ ,
\nonumber
\end{eqnarray}
where $d_H\left( i, p^k \right)$ denotes the Hamming distance bewteen the input
$i$ and the stored pattern $p^k$. 

In the final step I restore the memory gate to the state 
$|m\rangle$ by applying
the inverse transformation to eq. (\ref{ae}) and I 
apply the Hadamard gate  
to the control qbit $c_1$, thereby obtaining
\begin{eqnarray}
|\psi _4\rangle &&= H_{c_1} \prod_{j=n}^1 XOR_{i_j m_j}  
\ NOT_{m_j} \ |\psi_3\rangle \ ,
\label{ai} \\
|\psi_4\rangle &&= {1\over \sqrt{p}} \sum_{k=1}^p {\rm cos}\  {\pi \over
2n} d_H\left( i, p^k\right) 
|i; p^k; 0_1,\dots ,0_b\rangle 
\nonumber \\
&&+ {1\over \sqrt{p}} \sum_{k=1}^p {\rm sin}\  {\pi \over 2n}
d_H\left( i, p^k\right) 
|i; p^k; 1_1,\dots ,0_b\rangle .
\nonumber
\end{eqnarray}

The idea is now to repeat the above operations sequentially for all $b$ control
qbits $c_1$ to $c_b$. This gives 
\begin{eqnarray}
|\psi_{\rm fin}\rangle &&= {1\over \sqrt{p}} \sum_{k=1}^p \sum_{l=0}^b
\ {\rm cos}^{b-l} \left( {\pi\over 2n} d_H\left( i, p^k \right)\right) \times 
\nonumber \\
&&{\rm sin}^l \left( {\pi\over 2n} d_H\left( i, p^k \right)\right) 
\ \sum_{\left\{ J^l \right\}} |i; p^k; J^l\rangle ,
\label{al}
\end{eqnarray}
where $\left\{ J^l \right\}$ denotes the set of all binary numbers of
$b$ bits with exactly $l$ bits 1 and $(b-l)$ bits 0.

As in the case of the storing algorithm, there is a version of the information
retrieval algorithm in which the input is not loaded into an auxiliary quantum
register but rather into a unitary operator. Indeed, the auxiliary quantum
register is needed only by the operator (\ref{ae}) leading from (\ref{ad}) to
(\ref{af}). The same result (apart from an irrelevant overall sign) can be
obtained by applying
\begin{eqnarray}
I &&= \prod_{j=1}^n U_j \ ,
\nonumber \\
U_j &&= {\rm sin}\left( {\pi \over 2} i_j \right) 1 + i\ {\rm cos}
\left( {\pi\over 2} i_j \right) \sigma_2 \ ,
\label{newg}
\end{eqnarray}
directly on the memory state $|m\rangle$.  The rest of the algorithm is
the same, apart the reversing of the operator (\ref{ae}) which needs now the
operator $I^{-1}$. 

The end effect of the information retrieval algorithm represents
a rotation of the memory quantum state in the enlarged Hilbert
space obtained by adding $b$ control qbits.
Note that the overall effect of this rotation
is an overall amplitude concentration on memory states
similar to the input if there is a large number of $|0\rangle$ control qbits and
an amplitude concentration on states different to the input if there is a large
number of $|1\rangle$ control qbits. 
As a consequence, the most interesting
state for information retrieval purposes is the projection of $|\psi_{\rm
fin}\rangle$ onto the subspace with all control qbits in state $|0\rangle$.

There are two ways of obtaining this projection. The first is to repeat the
above algorithm and measure the control register
several times, until exactly the desired state for the control
register is obtained. If the number of such repetitions exceeds a preset
threshold $T$ the input is classified as "non-recognized" and the algorithm is
stopped. Otherwise, once $|c_1, \dots , c_b\rangle = |0_1, \dots, 0_b\rangle$ is
obtained, one proceeds to a measurement of the memory register $m$, which yields
the output pattern of the memory. 

The second method is to apply $T$ steps of the amplitude amplification algorithm
\cite{amam} rotating $|\psi_{\rm fin}\rangle$ towards its projection onto the
"good" subspace formed by the states with all control qbits in state
$|0\rangle$. To this end it is best to use the versions of the storing and
retrieving algorithms which do not need any auxiliary quantum register for
inputs. Let me define as $R(i)$ the input-dependent operator which rotates the
memory state in the Hilbert space enlarged by the $b$ control qbits towards the
final state $|\psi _{\rm fin}\rangle$ in eq. (\ref{al}) (where I now omit the
auxiliary register for the input): 
\begin{equation}
|\psi_{\rm fin}\rangle = R(i) \ |m;0_1,\dots,0_b\rangle \ .
\label{newh}
\end{equation}
By adding also the two utility qbits needed for the storing algorithm one can
then obtain $|\psi_{\rm fin}\rangle$ as a unitary transformation of the initial
state with all qbits in state $|0\rangle$:
\begin{equation}
|\psi_{\rm fin};00\rangle = R(i) M \ |0,\dots,0;0_1,\dots,0_b;00\rangle \ .
\label{newi}
\end{equation}
The amplitude amplification rotation of $|\psi_{\rm fin};00\rangle$ towards its
"good" subspace in which all $b$ control qbits are in state $|0\rangle$ is then
obtained \cite{amam} by repeated application of the operator 
\begin{equation}
Q = - R(i)MS_0 M^{-1}R^{-1}(i)S 
\label{newl}
\end{equation}
on the state $|\psi_{\rm fin};00\rangle$. Here $S$ conditionally changes the
sign of the amplitude of the "good" states with the $b$ control qbits in state
$|0\rangle$, while $S_0$ changes the sign of the amplitude if and only if the
state is the zero state $|0,\dots,0;0_1,\dots, 0_b;00\rangle$. 
As before, if a measurement of the control register after the
$T$ iterations of the amplitude amplification rotation yields $|0_1, \dots,
0_b\rangle$ one proceeds to a measurement of the memory register;
otherwise the input is classified as "non-recognized". 

Since the expected number of repetitions needed to measure the desired control
register state is $1/P_b^{\rm rec}$, with
\begin{equation}
P_b^{\rm rec} = {1\over p} \ \sum_{k=1}^p \ {\rm cos}^{2b} \left(
{\pi \over 2n} d_H \left( i; p^k\right) \right) 
\label{am}
\end{equation}
the probability of measuring $|c_1,\dots ,c_n\rangle = |0_1, \dots ,0_n\rangle$,
the threshold $T$ governs the {\it recognition efficiency} of the input
patterns. Note, however, that
amplitude amplification provides a quadratic boost \cite{amam} 
to the recognition efficiency since only $1/\sqrt{P_b^{\rm rec}}$ steps are
required to rotate $|\psi_{\rm fin}\rangle$ onto the desired subspace.
Accordingly, the threshold $T$ can be lowered to $\sqrt{T}$ with respect to the
method of projection by measurement. 

Once the input pattern $i$ is recognized, the measurement of the memory register
yields the stored pattern $p^k$ with probability
\begin{eqnarray}
P_b\left( p^k\right) &&= {1\over Z} \ {\rm cos}^{2b} \left( {\pi \over 2n}
d_H\left( i, p^k\right) \right) \ ,
\label{an} \\
Z &&= pP_b^{\rm rec} = \sum_{k=1}^p {\rm cos}^{2b} \left( {\pi \over 2n}
d_H\left( i, p^k\right) \right) \ .
\label{ao}
\end{eqnarray}
Clearly, this probability is peaked around those patterns which have the
smallest Hamming distance to the input. The highest probability of retrieval is
thus realized for that pattern which is most similar to the input. This is
always true, independently of the number of stored patterns. In particular there
are never spurious memories: the probability of obtaining as output a non-stored
pattern is always zero. As a consequence there are no 
restrictions on the loading factor $p/n$ coming from the information retrieval
algorithm.

In addition to the threshold $T$,
there is a second tunable parameter, namely the number $b$ of
control qbits. This new parameter $b$ controls the {\it identification
efficiency} of the quantum memory since, increasing $b$, the probability
distribution $P_b\left( p^k\right)$ becomes more and more peaked on the 
low $d_H\left( i, p^k \right) $ states, until
\begin{equation}
\lim_{b\to \infty} P_b\left( p^k \right) = \delta_{k k_{\rm min}}\ ,
\label{ap}
\end{equation}
where $k_{\rm min}$ is the index of the pattern (assumed unique for convenience)
with the smallest Hamming distance to the input.

The probability of recognition is 
determined by comparing (even) powers of cosines and
sines of the distances to the stored patterns. It is thus clear that the worst
case for recognition is the situation in which there is an isolated pattern,
with the remaining patterns forming a tight cluster spanning all the largest
distances to the first one. As a consequence,
the threshold needed to recognize all patterns
diminishes when the number of stored patterns
becomes very large, since, in this case, the distribution of patterns becomes
necessarily more homogeneous. Indeed, for the maximal number of stored patterns
$p=2^n$ one has $P_b^{\rm rec}= 1/2^b$ and the recognition efficiency
becomes also maximal, as it should be. 

While the recognition efficiency depends on comparing 
powers of cosines and sines of the same distances in the distribution, the
identification efficiency depends on comparing the (even) powers of
cosines of the different distances in the distribution.
Specifically, it is best when one of the distances is zero, while all others are
as large as possible, such that the probability of retrieval is completely
peaked on one pattern. As a consequence, the
identification efficiency is best when the recognition efficiency is worst 
and viceversa.  

The role of the parameter $b$ becomes familiar upon a closer examination of 
eq.( \ref{an}). Indeed, the quantum distribution described by this equation is
equivalent to a canonical Boltzmann distribution 
with (dimensionless) temperature
$t = 1/b$ and (dimensionless) energy levels 
\begin{equation}
E^k = -2 \ {\rm log} \ {\rm cos} \left( {\pi \over 2n} d_H\left( i, p^k
\right) \right) \ ,
\label{aq}
\end{equation}
with $Z$ playing the role of the partition function. 

The appearance of an effective thermal distribution suggests studying the
average behaviour of quantum associative memories via the corresponding
thermodynamic potentials. Before this can be done, however, one must deal with
the different distributions of stored patterns characterizing each individual
memory. To this end I propose to average also over this distribution, by keeping
as a tunable parameter only the minimal Hamming distance $d$ between the input
and the stored patterns. In doing so, one obtains an average description of the
average memory. This is essentially the replica trick used to derive the
behaviour of spin glasses \cite{parisi} and classical Hopfield models
\cite{neuralnetworks}.

As a first step it is useful to normalize the pattern representation 
by adding (modulo 2) to all
patterns, input included, the input pattern $i$. This clearly preserves all
Hamming distances and has the effect of normalizing the input to be the state
with all qbits in state $|0\rangle$. The Hamming distance $d_H\left( i, p^k
\right)$ becomes thus simply the number of qbits in pattern $p^k$ with value
$|1\rangle$. For loading factors $p/n \to 0$ in the limit $n\to \infty$
the partition function for the average memory takes 
then a particularly simple form:
\begin{equation}
Z_{\rm av} = {p\over N_{\lambda}} 
\ \sum_{\{\lambda \}} \ \sum_{j=d}^{n} \ \lambda_j
\ {\rm cos}^{2b}\left( {\pi \over 2} {j\over n}\right) \ ,
\label{ar}
\end{equation}
where $\lambda_j$ describes an unconstrained 
probability distribution such that $\sum_{j=d}^n
\lambda_j = 1$, $\{ \lambda \}$ is the set of such distributions and
$N_{\lambda}$ the corresponding normalization factor. For finite loading
factors, instead, the probabilities $\lambda_j$ become subject to constraints
which make things more complicated. 

I now introduce the free energy $F(b,d)$ by the usual definition
\begin{equation}
Z_{\rm av} = p \ {\rm e}^{-bF(b,d)} = Z_{\rm av}(b=0) \ {\rm e}^{-bF(b,d)} \ ,
\label{as}
\end{equation}
where I have chosen a normalization such that ${\rm exp}(-bF)$ describes the
deviation of the partition function from its value for $b=0$ (high effective
temperature). Since $Z/p$, and consequently also $Z_{\rm av}/p$ 
posses a finite, non-vanishing large-$n$ limit,
this normalization ensures that $F(b,d)$ is intensive, exactly like the energy
levels (\ref{aq}), and scales as a constant for large $n$. This is the only
difference with respect to the familiar situation in statistical mechanics.

The free energy describes the equilibrium of the system at effective temperature
$t=1/b$ and has the usual expression in terms of the internal energy $U$ and the
entropy $S$:
\begin{eqnarray}
F(t,d) &&= U(t,d) - tS(t,d) \ ,
\nonumber \\
U(t,d) && = \langle E \rangle _t \ ,\quad 
S(t,d) = {-\partial F(t,d) \over \partial t} \ .
\label{at}
\end{eqnarray}
Note that, with the normalization I have chosen in (\ref{as}), 
the entropy $S$ is
always a negative quantity describing the deviation from its maximal value 
$S_{\rm max} = 0$ at $t=\infty$.

By inverting eq.(\ref{aq}) with $F$ substituting $E$ one can also define an
effective (relative) input/output Hamming distance ${\cal D}$ 
at temperature $t$:
\begin{equation}
{\cal D}(t,d) = {2\over \pi} \ {\rm arccos} \ {\rm e}^{-F(t,d)\over 2} \ .
\label{au}
\end{equation}
This corresponds exactly to representing the recognition probability 
of the average memory as
\begin{equation}
\left( P_b^{\rm rec} \right) _{\rm av} = 
{\rm cos}^{2b} \left( {\pi \over 2} {\cal D}(b,d) \right) \ ,
\label{av}
\end{equation}
which can also be taken as the primary definition of the effective Hamming
distance. 

The function ${\cal D}(b,d)$ provides a complete description of the
behaviour of quantum associative memories in the limit $p/n \ll 1$. 
This can be used to tune their
performance. Indeed, suppose that one wants the memory to recognize and identify
inputs with up to $\epsilon n$ corrupted inputs with an efficiency of $\nu$
$(0\le \nu \le 1)$. Then one must choose a number $b$ of control qbits
sufficiently large that 
$\left( {\cal D}(b,\epsilon n) - \epsilon \right) \le 
\left( 1-\nu \right)$ and a
threshold $T$ of repetitions satisfying $T \ge 1/{\rm cos}^{2b} \left(
{\pi \over 2} {\cal D}(b,\epsilon n) \right) $, as illustrated in Fig. 1 below.

A first hint about the general behaviour of the effective distance function
${\cal D}(b,d)$ can be obtained by examining closer the energy eigenvalues
(\ref{aq}). For small Hamming distance to the input these reduce to
\begin{equation}
E^k \simeq {\pi^2\over 4} \left( {d_H\left( i, p^k\right) \over n}
\right) ^2\ ,\qquad {d_H\left( i, p^k \right) \over n} \ll 1\ .
\label{aw}
\end{equation}
Choosing again the normalization in which $|i\rangle = |0 \dots 0\rangle$ and
introducing a ``spin" $s_i^k$ with value $s_i^k = -1/2$ if qbit
$i$ in pattern $p^k$ has value $|0\rangle$ and $s_i^k=+1/2$ if qbit $i$ in
pattern $p^k$ has value $|1\rangle$, one can express the energy levels for 
$d_H/n \ll 1$ as
\begin{equation}
E^k = {\pi ^2\over 16} + {\pi ^2\over 4n^2}
\sum_{i,j} s_i^k s_j^k + {\pi ^2 \over 4n} \sum_i s_i^k \ .
\label{ay}
\end{equation}
Apart from a constant, this is the Hamiltonian of an infinite-range
antiferromagnetic Ising model in presence of a magnetic field. The
antiferromagnetic term favours configurations $k$ with half the spins up 
and half down, so that $s^k_{\rm tot} =\sum_i s^k_i = 0$, giving $E^k =
\pi^2/16$.The magnetic field, however, tends to align the
spins so that $s^k_{\rm tot} = -n/2$, giving $E^k = 0$. Since this is
lower than $\pi^2/16$, the ground state configuration is ferromagnetic, with all
qbits having value $|0\rangle$. At very low temperature (high $b$), where the
energy term dominates the free energy, one expects thus an ordered phase of the
quantum associative memory with ${\cal D}(t,d) = d/n$. This corresponds to a
perfect identification of the presented input. As the temperature is raised
($b$ decreased) however, the thermal energy embodied by the entropy term in the
free energy begins to counteract the magnetic field. At very high temperatures
(low $b$) the entropy approaches its maximal value $S(t=\infty) = 0$ (with the
normalization chosen here). If this value is approached faster than $1/t$, the
free energy will again be dominated by the internal energy . In this case,
however, this is not any more determined by the ground state but rather equally
distributed on all possible states, giving
\begin{eqnarray}
F(t=\infty) &&= U(t=\infty) = {-1\over 1-{d\over n}} 
\int_{d\over n}^1 \ dx
\ 2\ {\rm log} \ {\rm cos}\left( {\pi \over 2} x\right) 
\nonumber \\
&&= \left( 1+{d\over n}
\right) 2 \ {\rm log }2 + O\left( \left( {d\over n}\right) ^2\right) \ ,
\label{az}
\end{eqnarray}
and leading to an effective distance
\begin{equation}
{\cal D}(t=\infty, d) = {2\over 3} - {2\ {\rm log }2 \over \pi \sqrt{3}} 
\ {d\over n} + O\left( \left( {d\over n}\right) ^2\right) \ .
\label{azz}
\end{equation}
This value corresponds to a disordered phase with no correlation between input
and output of the memory. 

A numerical study of the thermodynamic potentials in (\ref{at}) and (\ref{au})
indeed confirms a phase transition from the ordered to the disordered phase as
the effective temperature is raised. In Fig. 1 I show the effective
distance ${\cal D}$ and the entropy $S$ 
for 1 Mb ($n=8 \times 10^6$) patterns and
$d/n= 1\%$ as a function of the inverse temperature $b$ (the entropy is rescaled
to the interval [0,1] for ease of presentation). At high temperature there is
indeed a disordered phase with $S=S_{\rm max} =0$ and ${\cal D} = 2/3$. At
low temperatures, instead, one is in the ordered phase with $S=S_{\rm min}$ and
${\cal D}=d/n=0.01$. The effective Hamming distance plays thus the role of the
order parameter for the phase transition. 

\begin{figure}
\centerline{\epsfysize=8.5cm\epsfbox{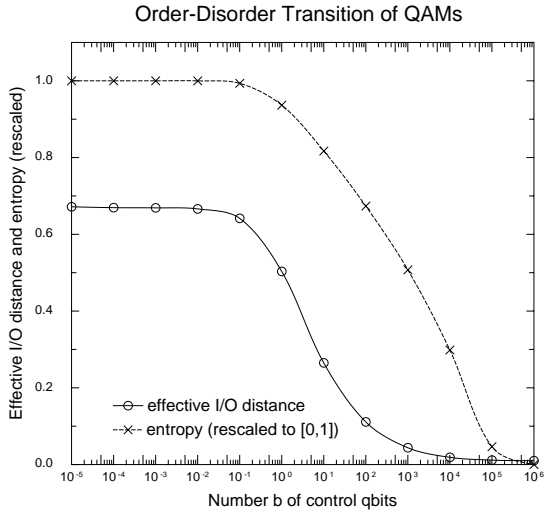}}
\caption{Effective input/output distance and entropy (rescaled to [0,1]) 
for 1Mb patterns and $d/n = 1\%$.}
\end{figure}

The phase transition occurs at $b_{\rm cr} \simeq 10^{-1}$. The physical regime
of the quantum associative memory ($b$ = positive integer) begins thus just
above this transition. For a good accuracy of pattern recognition one should
choose a temperature low enough to be well into the ordered phase.   

Having described at length the information retrieval mechanism for complete,
but possibly corrupted patterns, it is easy to incorporate also
incomplete ones. To this end assume that only $q<n$ qbits of the input are 
known and let me denote these by the indices $\{k1, \dots, kq\}$. After
assigning the remaining qbits randomly, there are two possibilities. One can
just treat the resulting complete input as a noisy one and proceed as above or,
better, one can limit the operator $\left( d_H \right)_m$ in the 
Hamiltonian (\ref{ag}) to
\begin{equation}
\left( d_H \right)_m = \sum_{i=1}^q \ \left( {\sigma_3+1\over 2} 
\right)_{m_{ki}} \ ,
\label{p}
\end{equation}
so that the Hamming distances to the stored patterns are computed on the
basis of the known qbits only. After this the pattern recall process continues
exactly as described above. This second possibility has the advantage that it
does not introduce random noise in the similarity measure but it has the
disadvantage that the operations of the memory have to be adjusted to the
inputs.

\section{Efficiency, complexity and memory tuning}
As anticipated in section 4, the effective i/o Hamming distance can be used to
tune the quantum associative memory to prescribed accuracy levels. Typically, it
is to be expected that increasing this accuracy will lead to an enhanced
complexity level. Before I even begin addressing this issue, however, I will
show that the information retrieval algorithm is efficient. 

First of all I would like to point out that, in addition to the standard
NOT, H (Hadamard), XOR, 2XOR (Toffoli) and nXOR gates
\cite{review} I have introduced 
only the two-qbit gates $CS^i$ in eq.
(\ref{addb}) and the unitary 
operator ${\rm exp}\left( i\pi {\cal H}/2n \right)$.
This latter can, however also be realized by simple gates
involving only one or two qbits. To this end I introduce the single-qbit gate
\begin{equation}
U = \left( \matrix{{\rm e}^{i{\pi\over 2n}}&0\cr
0&1\cr} \right) \ ,
\label{q}
\end{equation}
and the two-qbit controlled gate
\begin{equation}
CU^{-2} = |0\rangle \langle 0| \otimes 1 + |1\rangle \langle 1|
\otimes U^{-2} \ .
\label{r}
\end{equation}
It is then easy to check that ${\rm exp}\left( i\pi {\cal H}/2n \right)$ 
in eq. (\ref{af}) can be realized as follows:
\begin{equation}
{\rm e}^{i{\pi \over 2n} {\cal H}} \ |\psi _2\rangle = 
\prod_{i=1}^n \left( CU^{-2} \right) _{c m_i} \ \prod_{j=1}^n U_{m_j}
\ |\psi_2\rangle \ ,
\label{t}
\end{equation}
where $c$ is the control qbit for which one is currently repeating the
algorithm. Essentially, this
means that one implements first ${\rm exp}\left( i\pi  d_H /2n \right)$
and then one
corrects by implementing ${\rm exp}\left( -i\pi  d_H /n \right)$ on that part of
the quantum state for which the control qbit $|c\rangle$ is in state
$|1\rangle$. 

Using this representation for the Hamming distance operator one can count the
total number of simple gates that one must apply in order to implement one step
of the information retrieval algorithm. This is given by $(6n+2)$ using the
auxiliary register for the input and by $(4n+2)$ otherwise. This retrieval step
has then to be repeated for each of the $b$ control qbits. Therefore,
implementing the projection by repeated measurements, the overall complexity $C$
of information retrieval is bounded by 
\begin{equation}
C \le T b (6n+2) C_{\rm clon}\ ,
\label{neweqe}
\end{equation}
where $C_{\rm clon}$ is the complexity of the (probabilistic) cloning machine
that prepares a copy of the memory state.

The computation of the complexity is easier for the information retrieval
algorithm which uses the amplitude amplification technique. In this case the
initial memory is prepared only once by a product of the operators $M$, with
complexity $p(2n+3)+1$ and $R(i)$, with complexity $b(4n+2)$. Then one applies
$T$ times the operator $Q$, with complexity $p(4n+6) + b(8n+4) + 2 + C_S +
C_{S_0}$, where $C_S$ and $C_{S_0}$ are the polynomial 
complexities of the oracles implementing $S$ and $S_0$. This gives
\begin{eqnarray}
C &&= T\left[ p(4n+6) + b(8n+4) + 2 +C_S + C_{S_0}\right] +
\nonumber \\
&&+ p(2n+3) + b(4n+2)+1\ .
\label{newm}
\end{eqnarray}
As expected, this result depends on both $T$ and $b$, the parameters governing
the recognition and identification efficiencies. This represents exactly
the unavoidable tradeoff between accuracy and complexity. 

Suppose now one would like to recognize on average inputs with up to 1\% of
corrupted or missing bits and identify them with high accuracy. The effective
i/o Hamming distance ${\cal D}$ shown in Fig. 1 can then be used to determine
the values of the required parameters $T$ and $b$ needed to reach this accuracy
for the average memory with $p/n \ll 1$. 
For $b=10^4$ e.g., one has ${\cal D} = 0.018$, which
gives the average i/o distance (in percent of total qbits) if the minimum
possible i/o distance is 0.01. For this value of $b$ the recognition probability
is $1.5\ 10^{-4}$. With the measurement repetition technique one should thus set
the threshold $T=0.6 \ 10^4$. Using amplitude amplification, however, one needs
only around $T=80$ repetitions. 

I would like to conclude by stressing that the values of $b$ and $T$ obtained by
tuning the memory with the effective i/o Hamming distance
become $n$-independent 
for large values of $n$. This is because they are intensive variables unaffected
by this "thermodynamic limit". For any $p$ polynomial in $n$ the information
retrieval can then be implemented efficiently and the overall complexity is
determined by the accuracy requirements via the n-independent 
parameters $T$ and $b$. 

\section{Conclusion}
I would like to conclude this review by stressing the reason why a quantum
associative memory works better than its classical counterpart.

In classical associative memories, the information about the patterns to recall
is typically stored in an energy functional. When retrieving information, 
the input configuration evolves to the corresponding output, driven by the
memory functional.
The capacity shortage is due to a phase transition in the statistical
ensemble governed by the memory energy functional. Spurious memories, i.e. 
spurious metastable minima
not associated with any of the original patterns become important for loading
factors $p/n > 0.14$ and wash out completely the memory. So, in the low $p/n$
phase the memory works perfectly in the sense that it outputs always the stored
pattern which is most similar to the input. For $p/n > 0.14$, instead, there is
an abrupt transition to total amnesia caused by spurious memories. 

Quantum associative memories work better than classical ones since they are free
from spurious memories. The easiest way to see this is in the formulation
\begin{equation}
|m\rangle = M \ |0\rangle \ .
\label{newn}
\end{equation}
All the information about the stored patterns is encoded in the unitary
operator $M$. This is such that all quantum states that do not correspond to
stored patterns have exactly vanishing amplitudes. 

An analogy with the classical Hopfield model \cite{neuralnetworks} can be
established as follows. Instead of generating the memory state from the initial
zero state one can start from a uniform superposition of the computational
basis. This is achieved by the operator $MW$ defined by 
\begin{eqnarray}
|m\rangle &&= M W  \ {1\over \sqrt{2^n}} \sum_{j=0}^{2^n-1} |j\rangle \ ,
\nonumber \\
W &&\equiv \prod_{j=1}^n H_j \ .
\label{newo}
\end{eqnarray}
Now, the same result can also be obtained by Grover's algorithm, 
or better by its generalization with zero failure rate \cite{long}. Here the
state $|m\rangle$ is obtained by applying to the uniform superposition of the
computational basis q times the search operator $X$ defined by
\begin{eqnarray}
|m\rangle &&= X^q \ {1\over \sqrt{2^n}} \sum_{j=0}^{2^n-1} |j\rangle \ ,
\nonumber \\
X &&\equiv -W J_0 W J \ ,
\label{newp}
\end{eqnarray}
where $J$ rotates the amplitudes of the states corresponding to the patterns 
to be stored by a phase $\phi$ which is very close to $\pi$ (the original Grover
value) for large $n$ and $J_0$ does the same on the zero state. 
Via the two equations (\ref{newo}) and (\ref{newp}), the memory operator $M$
provides an implicit realization of the phase shift operator $J$. Being a
unitary operator, this can always be written as an exponential of an hermitian
Hamiltonian ${\cal H}$, which is the quantum generalization of a classical
energy functional. By defining $J\equiv {\rm exp}(-i{\cal H})$ one obtains an
energy operator which is diagonal in the computational basis. The energy
eigenvalues of this operator are such that the patterns to be stored have energy
$E=-\phi \simeq -\pi$ while all others have energy $E=0$. 
This formulation is the exact quantum
generalization of the Hopfield model; the important point is that the operator
$M$ realizes efficiently a dynamics in which the patterns to be stored are
always, for all numbers $p$ of patterns, 
the exact global minima of a quantum energy landscape, without the
appearance of any spurious memories. 
The price to pay is the probabilistic nature of the information retrieval
mechanism.
As always in quantum mechanics, the dynamics determines only the evolution of
probability distributions and the probabilistic aspect is brought in by the
collapse of this probability distributions upon measurement. Therefore,
contrary to the classical Hopfield model in the low $p/n$ phase, 
one does not always have the absolute guarantee that an input 
is recognized, and identified correctly as the
stored pattern most similar to the input, even if this state has the highest
probability of being measured.
But, after all, the same happens also to the human brain.

\end{document}